\documentclass{elsart}
\usepackage{amssymb,amsmath}
\usepackage {graphicx}
\usepackage[english]{babel}

\newcommand{\bigfig}[1]{\centering{\includegraphics[width=0.9\textwidth]{#1}}}
\newcommand{\onefig}[1]{\centering{\includegraphics[width=0.618\textwidth]{#1}}}
\newcommand{\twofig}[1]{\includegraphics[width=0.48\textwidth]{#1}}

\renewcommand{\Re}{\mathop{\mathrm{Re}}}
\renewcommand{\Im}{\mathop{\mathrm{Im}}}
\newcommand{\matr}[1]{\mathrm{\mathbf{#1}}}
\renewcommand{\vec}[1]{\mathbf{#1}}
\newcommand{\intern}{\mathop{\mathrm{Int}}}

\newtheorem{theorem}{Theorem}

\begin{document}

\begin{frontmatter}

\title{Hyperbolic attractor of Smale-Williams type in a system of two
coupled non-autonomous amplitude equations}

\author[SGAP]{Pavel V. Kuptsov\corauthref{cor}}\ead{p.kuptsov@rambler.ru}
\author[IRE]{Sergey P. Kuznetsov}
\author[IRE]{Igor R. Sataev}

\address[SGAP]{Department of Informatics, Saratov State Law Academy,
Volskaya 1, Saratov 410056, Russia}

\address[IRE]{Kotel'nikov Institute of Radio Engineering and Electronics of RAS,
Saratov Branch, Zelenaya 38, Saratov 410019, Russia}

\corauth[cor]{Corresponding author.}

\begin{abstract}
Recently, a system with uniformly hyperbolic attractor of
Smale-Williams type has been suggested [Kuznetsov, Phys. Rev.
Lett., 95, 144101, 2005]. This system consists of two coupled
non-autonomous van der Pol oscillators and admits simple physical
realization. In present paper we introduce amplitude equations for
this system and prove that the attractor of the system of
amplitude equations is also uniformly hyperbolic. Also we
represent qualitative illustrations as well as quantitative
characteristics of a chaotic dynamics on this attractor.
\end{abstract}


\end{frontmatter}

\section{Introduction}

One of the interesting problems in nonlinear dynamics is a
development of physical applications of the hyperbolic theory. A
trajectory in phase space of a dynamical system is hyperbolic when
a tangent vector space on each its point can be described as a
direct sum of subspaces spanned on stable and unstable
perturbation vectors and this representation is invariant along
the trajectory. Dissipative system with attractor containing only
hyperbolic trajectories demonstrates strong chaotic properties and
permits advanced mathematical analysis. Monographs and textbooks
on nonlinear dynamics provide examples of such attractors, but
most of them are artificial mathematical constructions like Plykin
attractor and Smale-Williams
solenoid~\cite{Sinai79,EckRuel85,Devaney89,Shil97,KatHass95,AfrHsu03,Ott93,AnAst02}.
On the other hand, attractors of realistic systems with
complicated dynamics, like the Lorenz
model~\cite{AfrByk77,MischMroz95}, do not relate to the class of
uniformly hyperbolic, so the hyperbolic theory can not be applied
to them in corpore.

In a recent paper~\cite{Hyp} one of the authors suggested an
implementation of the Smale-Williams attractor in a system of two
coupled non-autonomous van der Pol oscillators with natural
frequencies relating as 1:2. Subsystems are activated alternately
due to a slow variation of their excitation parameters by an
external force. The excitation is passed from one oscillator to
another, so that a phase of the oscillations is doubled within
each full cycle. This system was realized as an electronic devise
and studied experimentally~\cite{HypExper}. Assumption about the
hyperbolicity of the attractor was based on the fact that a phase
of one of the oscillators, being measured at successive stages of
its excitation, was obeyed to the Bernoulli map, as must be for
the Smale-Williams attractor. Moreover, computations revealed a
robustness of the attractor: Its Cantor-like transverse structure
and positive Lyapunov exponent were insensitive to variations of
parameters in the equations. Further analysis~\cite{KuzSat07}
showed that 4D phase space of a Poincar\'{e} map, describing a
successive states of the system over a period of the external
forcing, contains a toroidal absorbing domain (a direct product of
1D circle and 3D ball) that is mapped into itself according to the
Smale-Williams procedure. Performed numerical analysis proved the
validity of sufficient conditions of the hyperbolicity that are
formulated in terms of expanding and contracting cones in tangent
vector space.

An idea of alternate excitation of two oscillators that pass the
excitation from one to another can be applied to amplitude
equations which are derived for the van der Pol equations via the
method of slow varying amplitudes. Actually, equations of this
type can be obtained for a wide variety of dynamical systems
because, in fact, they correspond to a normal form of
Andronov-Hopf bifurcation of birth of a limit cycle. For example,
amplitude equations are employed in Landau's theory of
turbulence~\cite{Landau1,Landau2}.

Reasoning in the manner of Landau's theory, we can assume that a
spatially extended system have two modes with frequencies relating
as 1:2. Some additional periodic and slow component acts on
these modes so that they excite alternately and pass the
excitation from one to another through a nonlinear coupling. Then,
according to the idea of~\cite{Hyp}, this can result in the
formation of a hyperbolic attractor of Smale-Williams type.

This paper is devoted to a study of a system of two coupled
non-autonomous amplitude equations. We provide a proof of
hyperbolicity of an attractor of this system and also discuss some
attributes of its hyperbolic dynamics.

\section{System of two coupled non-autonomous van der Pol
oscillators and corresponding amplitude equations}

The starting point of our analysis is a system of two coupled van
der Pol oscillators that has been proposed in~\cite{Hyp}:
\begin{equation}
  \label{eOrigVDPSyst}
  \begin{gathered}
    \ddot{x}-[A\cos (2\pi \tau/P)-x^2]\dot{x}+\omega_0^2 x = \varepsilon y \cos\omega_0 \tau,\\
    \ddot{y}-[-A\cos (2\pi \tau/P)-y^2]\dot{y}+4\omega_0^2 y = \varepsilon x^2,
  \end{gathered}
\end{equation}
where $\tau$ denotes time. The oscillators have natural
frequencies $\omega_0$ and $2\omega_0$, respectively. Their
bifurcation parameters controlling a birth of a limit cycle
undergo a variation with a period $P$ and amplitude $A$. The
variations are in counter phases: when the first oscillator is
excited the second one is not and vice versa. The forcing is
supposed to be slow, i.e., its period is much larger than natural
periods of the oscillators. The oscillators are coupled and the
intensity of this interaction is controlled by a parameter
$\varepsilon$.

The subject of our study shall be a system of amplitude equations
corresponding to Eqs~\eqref{eOrigVDPSyst}. To derive these
equation we follow the standard technic and assume a solution to
Eqs~\eqref{eOrigVDPSyst} to be oscillations with frequencies
$\omega_0$ and $2\omega_0$, respectively, and with slow varying
complex amplitudes:
\begin{equation}
  \label{eAnsatz}
  \begin{gathered}
    x(\tau)=a(\tau)e^{i\omega_0 \tau}+a^*(\tau)e^{-i\omega_0 \tau},\\
    y(\tau)=b(\tau)e^{2i\omega_0 \tau}+b^*(\tau)e^{-2i\omega_0 \tau}.
  \end{gathered}
\end{equation}
Upper index ``*'' means complex conjugation. The derivatives of
complex amplitudes $a$ and $b$ should satisfy additional
conditions
\begin{equation}
  \begin{gathered}
    \dot{a}e^{i\omega_0 \tau}+\dot{a}^*e^{-i\omega_0 \tau}=0, \\
    \dot{b}e^{2i\omega_0 \tau}+\dot{b}^*e^{-2i\omega_0 \tau}=0.
  \end{gathered}
\end{equation}
Taking these conditions into account we substitute~\eqref{eAnsatz}
into~\eqref{eOrigVDPSyst} and average over the period
$2\pi/\omega_0$. Because the complex amplitudes $a$ and $b$ are
supposed to be slow, the resulting equations have the form:
\begin{equation}
  \begin{gathered}
    \dot{a} =   (A/2) a \cos (2\pi \tau/P) - |a|^2 a/2 - i \varepsilon b   / (4\omega_0), \\
    \dot{b} = - (A/2) b \cos (2\pi \tau/P) - |b|^2 b/2 - i \varepsilon a^2 / (4\omega_0).
  \end{gathered}
\end{equation}
These equations allows the following rescaling of time variable
and parameters:
\begin{equation}
  \label{eRescale}
  t=\tau/2,\; T=P/2, \;\epsilon=\varepsilon/2\omega_0.
\end{equation}
As a result, we obtain a sought system of amplitude equations:
\begin{equation}
  \label{eAmplEqns}
  \begin{gathered}
    \dot{a} =   A a \cos (2\pi t/T) - |a|^2 a - i \epsilon \, b, \\
    \dot{b} = - A b \cos (2\pi t/T) - |b|^2 b - i \epsilon \, a^2.
  \end{gathered}
\end{equation}

To clarify the nature of dynamics of the system~\eqref{eAmplEqns},
we consider a behavior of amplitudes and phases of complex
variables $a$ and $b$. Define phases within the interval
$[0,2\pi)$: $\phi = \arg a$, $\psi = \arg b$. Suppose that the
first oscillator is excited and its amplitude $|a|$ is high.
Hence, the second one is suppressed so its amplitude $|b|$ is
small. The coefficients in Eqs.~\eqref{eAmplEqns} are real except
the coupling. It means that the phases can be changed only as a
result of interaction between subsystems. But when $a$ is excited
$|b|$ is small and its action on $a$ is negligible. Thus, the
phase of $a$ remains constant during the excitation stage. The
opposite influence from $a$ to $b$ is high and the coupling term
is proportional to $a^2$. It means that after a half period $T/2$
at the threshold of its own excitation the oscillator $b$ inherits
a doubled phase of $a$ (the phase also gets a shift $-\pi/2$
because of the imaginary unit at the coupling term). Now the roles
of the subsystems are exchanged. The phase of $b$ remains constant
when this subsystem is excited and at the end, after the other
half period $T/2$, the phase is returned back to $a$ through a
linear coupling term (also with the shift $-\pi/2$). As a result,
the first oscillator $a$ doubles its phase during the period~$T$.

\begin{figure}
  \centering a)\twofig{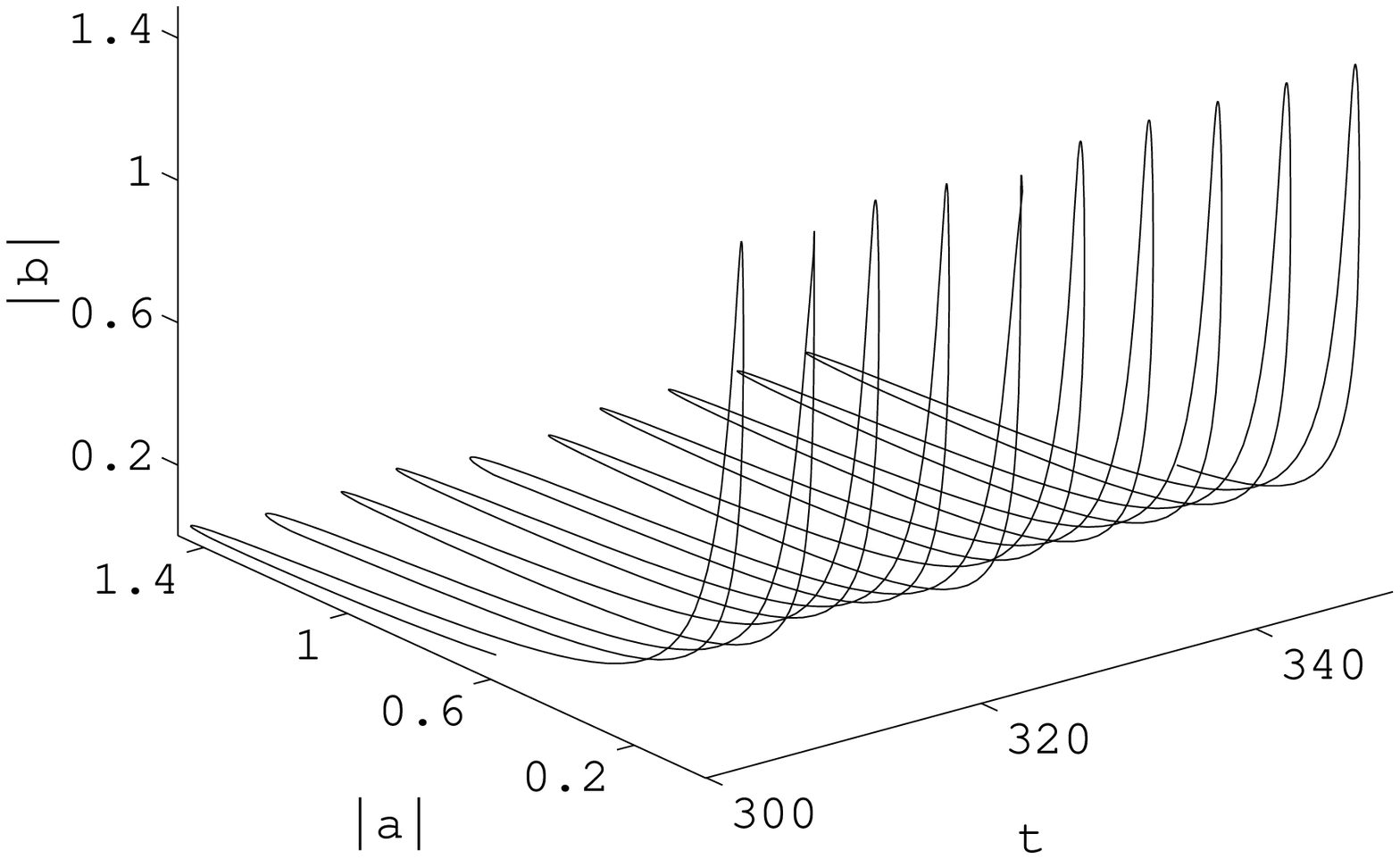}b)\twofig{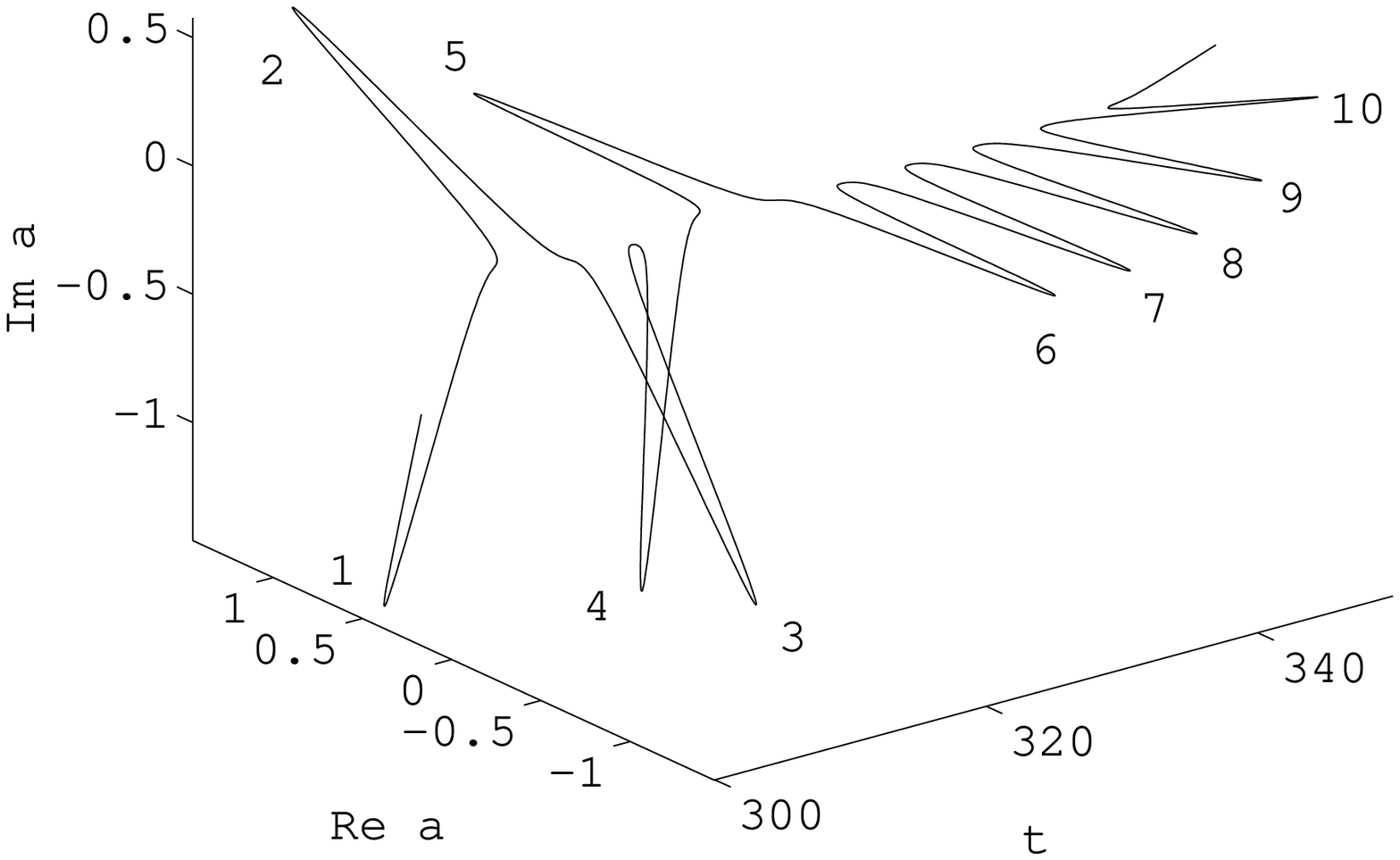}
  \caption{\label{fTimSer}Illustration of a dynamics of the
  system~\eqref{eAmplEqns}: alternate excitation and decay of the
  components $a$ and $b$ (panel (a)) and time dependance of
  real and imaginary parts of $a$ (panel (b)). Observe that
  being excited the component $a$ has almost constant phase,
  but the phase is doubled at every new stage of the excitation
  (compare, for example, spikes labelled $4$, $5$ and $6$).}
\end{figure}

Fig.~\ref{fTimSer} demonstrates a numerical solution to
Eqs.~\eqref{eAmplEqns} illustrating these considerations. Here and
below the solutions are obtained via Runge-Kutta method of the
fifth order with a step-size control~\cite{Recipes}. Panel (a)
demonstrates an alternate excitation of the subsystems $a$ and
$b$. Notice, that each time the amplitude $|a|$ attains its
maximum, the height of the maximum is a bit different from the
previous one. The same is for the amplitude $|b|$. This is a
manifestation of a chaotic nature of the discussed dynamics. Panel
(b) represents a time dependance of a complex variable $a$. We
observe a series of spikes corresponding to the stages of the
excitation of $a$. Projections of spikes on the plane $\{\Re a$,
$\Im a\}$ are almost straight lines. This confirms our previous
conclusion that the phase of an excited subsystem, i.e., an angle
between the projection and the real axis $\Re a$, remains constant
or, at least, almost constant. This panel also illustrates the
phase doubling after each period $T$: compare angles between
spikes $4$, $5$ and $6$.

The above discussion allows us to write down a map for a series of
phases $\phi_n = \arg a(nT)$ that are measured over the time step
$T$:
\begin{equation}
  \label{eBernulMap}
  \phi_{n+1}=2\phi_n - \pi  \mod 2\pi.
\end{equation}
Up to a constant addition (that can be eliminated by a shift of
the origin of the phase) this map coincides with the well known
Bernoulli map~\cite{ShusChaos}. It demonstrates a chaotic dynamics
and the chaos is homogeneous: a rate of exponential divergence of
two close trajectories is identical in each point of the phase
space and equal to $\ln 2$.

\begin{figure}
  \centering \twofig{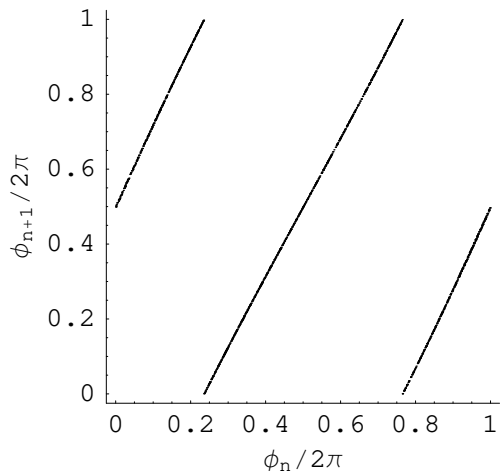}
  \caption{\label{fBernul}Numerically computed map for the phase
  $\phi_n=\arg a(nT)$ measured over the time step $T$ at
  $A=3$, $T=5$ and $\epsilon=0.05$.}
\end{figure}

Fig.~\ref{fBernul} show $\phi_{n+1}=\arg a(nT+T)$ vs. $\phi_n=\arg
a(nT)$ computed numerically for the system~\eqref{eAmplEqns}. One
can see a very high correspondence with the
map~\eqref{eBernulMap}. The most important point is the
topological equivalence between the empirical map and the
Bernoulli map: one full circle that passes a preimage (i.e., turn
of a phase on $2\pi$) implies two passes of a circle for an image
($4\pi$ turn of a phase).

The correspondence with the Bernoulli map presumes that among
Lyapunov exponents of the system~\eqref{eAmplEqns} the one should
be equal to $\ln 2/T$. To compute the Lyapunov exponents we employ
the Benettin's algorithm~\cite{BenettinLyap1,ParkerChua} that
requires to solve simultaneously Eqs.~\eqref{eAmplEqns} and four
sets of linearized equations for small perturbations:
\begin{equation}
  \label{eAmplEqnsVariat}
  \begin{gathered}
    \delta \dot{x}_0= A\cos(2\pi t/T){\delta x_0}-(3x_0^2+x_1^2){\delta x_0}-2x_0 x_1 {\delta x_1}+ \epsilon{\delta x_3}, \\
    \delta \dot{x}_1= A\cos(2\pi t/T){\delta x_1}-(x_0^2+3x_1^2){\delta x_1}-2x_0 x_1 {\delta x_0}- \epsilon{\delta x_2}, \\
    \delta \dot{x}_2=-A\cos(2\pi t/T){\delta x_2}-(3x_2^2+x_3^2){\delta x_2}-2x_2 x_3 {\delta x_3}+2\epsilon(x_0 {\delta x_1}+x_1 {\delta x_0}), \\
    \delta \dot{x}_3=-A\cos(2\pi t/T){\delta x_3}-(x_2^2+3x_3^2){\delta x_3}-2x_2 x_3 {\delta x_2}-2\epsilon(x_0 {\delta x_0}-x_1 {\delta x_1}),
  \end{gathered}
\end{equation}
where $x_0=\Re a$, $x_1=\Im a$, $x_2=\Re b$, $x_3=\Im b$, and
$\delta x_0$, $\delta x_1$, $\delta x_2$,  $\delta x_3$ denotes
small perturbations to these values. Before the start of the
computations, we initialize each set of linearized equations by a
unit perturbation vector, so that four these vectors comprise an
orthogonal system. Advancing the solution, we perform the
Gram-Schmidt orthogonalization~\cite{ParkerChua} of these vectors
after each time interval $T$ and normalize them to prevent a
numerical overflow. Four Lyapunov exponents $\lambda_k$ appear as
time averaged logarithms of norms of the perturbation vectors.
Finally, we multiply the computed values on $T$, to obtain the
Lyapunov exponents for corresponding map, describing the dynamics
of the system with time step $T$: $\Lambda_k=\lambda_k T$

\begin{figure}
  \onefig{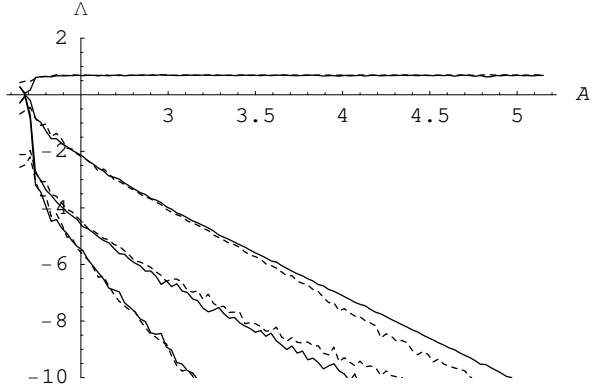}
  \caption{\label{fLyap}Lyapunov exponents $\Lambda_k$ vs. parameter $A$
  for the system~\eqref{eAmplEqns} at $T=5$
  and $\epsilon=0.05$. Dash lines indicate corresponding Lyapunov exponents
  for the initial system~\eqref{eOrigVDPSyst} at $P=10$, $\omega_0=2\pi$ and
  $\varepsilon=0.2\pi$.}
\end{figure}

Fig.~\ref{fLyap} demonstrates the exponents $\Lambda_k$ as
functions of the parameter $A$. The largest exponent is positive
for a wide range of parameter values. This is the evidence of a
chaotic nature of the observed dynamics. Three others are always
negative. For example, at $A=3$ the Lyapunov exponents are
$\Lambda_1\approx 0.691$, $\Lambda_2\approx -4.06$,
$\Lambda_3\approx -6.48$, and $\Lambda_4\approx -9.06$. As typical
for maps and non-autonomous continuous time systems, a zero
exponent is absent. The largest exponent is close to $\ln 2$. This
corresponds to our qualitative conclusion that the dynamics of the
phase of $a$ is obeyed, to some degree of accuracy, by the
Bernoulli map~\eqref{eBernulMap}. Also notice that the largest
exponent is almost independent on $A$, and the others vary rather
smoothly, without peaks and canyons. This can be treated as a
manifestation of a robustness of the observed chaotic oscillations
which, in turn, indicates a hyperbolic nature of an underlying
chaotic attractor.

This behavior of Lyapunov exponents agrees very well with the
behavior of the exponents of the initial system of two van der Pol
oscillators~\eqref{eOrigVDPSyst}~\cite{Hyp,HypExper}. In
Fig.~\ref{fLyap} the Lyapunov exponents of the initial system are
plotted by dashed lines. Values of parameters, being recalculated
as required by Eq.~\eqref{eRescale}, correspond to the parameters
of the system~\eqref{eAmplEqns}. The largest Lyapunov exponents of
two systems are almost identical and three others pairs are also
close to each other. We treat this good correspondence between the
initial system and the approximate amplitude equation as one more
demonstration of a robustness. In turn, the robustness is related
to a hyperbolic nature of the observed dynamics.

\section{\label{sPogl}Stroboscopic Poincar\'{e} map and absorbing domain}

Consider our system~\eqref{eAmplEqns} in a certain time moment
$t_n=nT$. Its instantaneous state is given by a vector
$\vec{x}_n=\{\Re a(t_n), \Im a(t_n), \Re b(t_n), \Im b(t_n)\}$. If
we take this vector as an initial state and integrate the
equations over the period $T$, we find a new vector
$\vec{x}_{n+1}=\{\Re a(t_{n+1}), \Im a(t_{n+1}), \Re b(t_{n+1}),
\Im b(t_{n+1})\}$. This vector is unambiguously determined by the
vector $\vec{x}_n$. Hence, we can define a mapping that operates
in $\mathbb{R}^4$:
\begin{equation}
  \label{eMap}
  \vec{x}_{n+1}=\matr{T}(\vec{x}_n).
\end{equation}
This equation defines a stroboscopic Poincar\'{e} map for the
system~\eqref{eAmplEqns}. From a geometric point of view, each
vector $\vec{x}_n$ belongs to a 4D hyperplane $t=t_n=nT$ which is
a section of a flow of trajectories in the 5D extended phase space
$\{\Re a, \Im a, \Re b, \Im b, t\}$. Because of periodicity of the
phase in $t$, these hyperplanes can be identified. So, we can say
that Eq.~\eqref{eMap} maps 4D hyperplane $\{\Re a, \Im a, \Re b,
\Im b\}$ onto itself.

The map $\matr{T}$ is generated by differential equations whose
right-hand sides are smooth and limited within a finite region of
$a$ and $b$. According to the existence, uniqueness, continuity
and differentiability theorems, the map $\matr{T}$ is one-to-one
diffeomorphism belonging to class $C^{\infty}$)~\cite{Arnold92}.

In course of iterations, the map
$\vec{x}_{n+1}=\matr{T}(\vec{x}_n)$ expands a volume element in a
direction, associated with the phase $\phi$, see
Eq.~\eqref{eBernulMap}, while three others directions are
contracting. Taking into account a periodicity of the phase, we
can consider a 4D toroid which is a direct product of a 1D circle
and 3D ball. One iteration of the map corresponds to a
longitudinal stretch and transverse contraction of this object
followed by its kinking and insertion into the initial area as a
double loop. This exactly corresponds to the Smale and Williams
procedure except that our map is embedded into 4D rather then 3D
state space.

Mentioned toroid with its interior is an absorbing domain which we
denote as $U$. Points, belonging to this area, fall into the
interior of $U$ under the action of the map $\matr{T}$:
$\matr{T}(U)\subset \intern U$. An attractor may be defined as an
intersection of all images of $U$, generated by successive
iterations of $\matr{T}$: $A=\bigcap_{\,n=1}^{\,\infty}
\matr{T}^n(U)$.

Consider the space $\{\Re a, \Im a, \Re b, \Im b\}$ and define a
torus as
\begin{equation}
  \label{eTorus}
  (|a|-r)^2+|b|^2=(\chi d)^2.
\end{equation}
Here $r$ denotes a radius of the 1D circle, and $\chi d$ is a
radius of the 3D ball that appears in a radial section of the
torus by a hyperplane perpendicular to the circle. Parameter
$\chi$ is introduced to scan points both on the surface ($\chi=1$)
and inside the absorbing domain, $0\leq \chi\leq 1$. The
torus~\eqref{eTorus} can also be defined parametrically via the
equations
\begin{equation}
  \label{eTorusParam}
  \begin{gathered}
    \Re a=(\chi d\cos\theta+r)\cos\phi, \; \Re b=\chi d\sin\theta\cos\psi, \\
    \Im a=(\chi d\cos\theta+r)\sin\phi, \; \Im b=\chi d\sin\theta\sin\psi.
  \end{gathered}
\end{equation}
Here parameters $\phi$ and $\psi$ can be treated as phases of $a$
and $b$, respectively.

Numerical values of $r$ and $d$ should be found for the attractor
to be fitted inside the torus. Consider specific values of the
parameters of the system, say, $A=3$, $T=5$, $\epsilon=0.05$, and
compute a large amount of points belonging to the attractor of the
map~\eqref{eMap}. Projections of these points to the plane $\{\Re
a, \Im a\}$ fall somewhere in a vicinity of a circle whose radius
$r$ can be found as a mean amplitude $|a|$ of these points. For
our specific parameters we find $r=0.631166$.

\begin{figure}
  \onefig{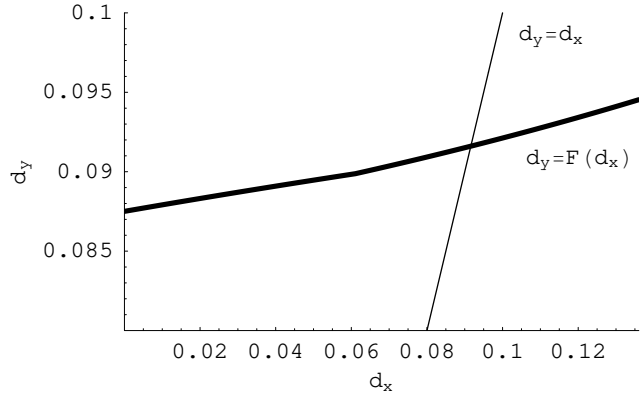}
  \caption{\label{fFindAbsorb}Selection of the parameters of an
  absorbing domain (see explanations in the text).}
\end{figure}

To compute $d$, let us consider a series of tori parameterized by
the section radius $d_x$ that grows from $0$ with some small step
(we assume here that $\chi=1$). All the tori have identical
radiuses $r$. For each $d_x$ we cover the surface of a
corresponding torus by a mesh, using parametric
Eqs.~\eqref{eTorusParam}. In our computations the angle variables
$\phi$, $\psi$ and $\theta$ were varied with the step $2\pi/50$.
The map~\eqref{eMap} is iterated once form each node of the mesh.
An image point falls on the surface of a new torus whose section
radius is equal to $\sqrt{(|a|-r)^2+|b|^2}$, where $|a|$ and $|b|$
are related here to the image point. Thus, the torus $d_x$
generates a set of ``image'' tori. Because we are seeking for an
absorbing domain, we must take the worst case, i.e., the image
torus with the largest section radius. Denoting this radius as
$d_y$, we obtain a function $d_y=F(d_x)$ that is plotted in
Fig.~\ref{fFindAbsorb}. This function monotonously grows and meets
the line $d_y=d_x$ at $d_x\approx 0.0918$. All tori to the right
from this point can be taken as an absorbing domain because their
surfaces, as well as the interior, are mapped into their interior.
We take a torus whose section radius $d$ is about $1\%$ above the
intersection point. So, the set of parameters of the system and
corresponding parameters of the absorbing domain are the
following:
\begin{equation}
  \label{eNumVal}
  \begin{gathered}
    A=3, \; T=5, \; \epsilon=0.05, \\
    r=0.631166, \; d=0.0927.
  \end{gathered}
\end{equation}

\begin{figure}
  \centering \bigfig{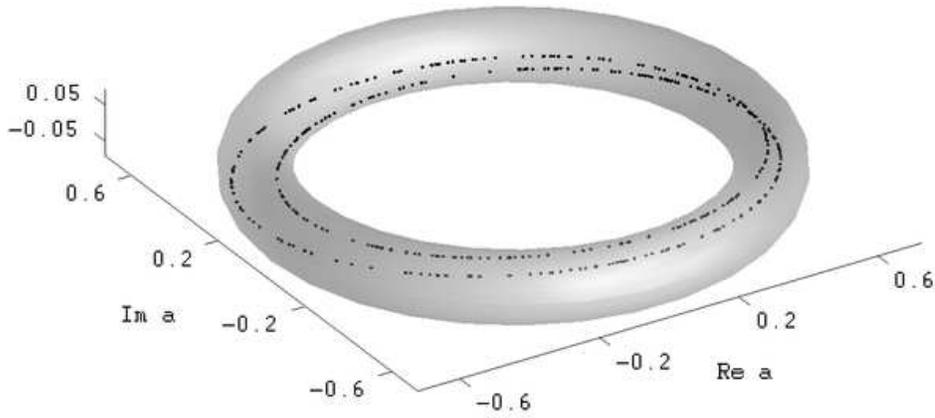}
  \caption{\label{fAbsorbAttr}3D projection of a toroidal absorbing
  domain and enclosed attractor of the Poincar\'{e}
  map~\eqref{eMap}.}
\end{figure}

Fig.~\ref{fAbsorbAttr} demonstrates 3D view of the absorbing
domain with an enclosed attractor. Points of the attractor are
project from 4D space $\{\Re a,\Im a, \Re b, \allowbreak \Im b\}$
along axis $\Im b$ onto the section plane $\psi=0$.

\section{Sufficient condition for hyperbolicity:
general formulation of a method}

To verify hyperbolic nature of the attractor of the
system~\eqref{eAmplEqns}, we shall follow the method developed
in~\cite{KuzSat07} for the attractor of two non-autonomous van der
Pol oscillators~\eqref{eOrigVDPSyst}. In this section we reproduce
the details of this method.

The central point is the theorem on expanding and contracting
cones~\cite{Sinai79,KatHass95,Hunt2000}. Unlike the general case,
it is sufficient here to deal with a diffeomorphism of class
$C^\infty$ in the Euclidian space $\mathbb{R}^4$, namely, we
consider the Poincar\'{e} map $\matr{T}(\vec{x})$. Let
$\matr{DT}_x$ be the Jacobi matrix of the map
$\vec{x}^\prime=\matr{T}(\vec{x})$ at $\vec{x}$:
$\matr{DT}_x=\{\partial x_i^\prime/\partial x_j\}$, $i,j=0,1,2,3$,
and let $\matr{DT}_x^{-1}$ designates the derivative matrix for
the inverse map $\matr{T}^{-1}(x)$. Also, let $\delta\vec{x}$
denotes the vector of a small perturbation to $\vec{x}$. In a
linear approximation the evolution of a perturbed state
$\vec{x}+\delta\vec{x}$ corresponds to transformation of the
vector $\delta\vec{x}$ according to a linear mapping $\delta
\vec{x}^\prime= \matr{DT}_x\delta \vec{x}$. Vectors
$\delta\vec{x}$ form a tangent space $\mathbb{V}_x$ associated
with $\vec{x}$.

\begin{theorem}[\cite{Sinai79,KatHass95,Hunt2000}]
Suppose that a diffeomorphism $\matr{T}$ of class $C^\infty$ maps
a bounded domain $U\subset \mathbb{R}^4$ into itself:
$\matr{T}(U)\subset\intern U$, and $A\subset\intern U$ is an
invariant set for the diffeomorphism. The set $A$ will be
uniformly hyperbolic if there exists a constant $\gamma>1$ and the
following conditions hold:
\begin{enumerate}
\item The expanding and contracting cones $S_x^\gamma$ and $C_x^\gamma$ may
be defined in the tangent space $\mathbb{V}_x$ at each $\vec{x}\in
A$, such that $||\matr{DT}_x \vec{u}||\geq\gamma ||\vec{u}||$ for
all $\vec{u}\in S_x^\gamma$, and
$||\matr{DT}_x^{-1}\vec{v}||\geq\gamma||\vec{v}||$ for all
$\vec{v}\in C_x^\gamma$; moreover, for all $\vec{x}\in A$ they
satisfy $S_x^\gamma \cap C_x^\gamma=\varnothing$ and
$S_x^\gamma+C_x^\gamma=\mathbb{V}_x$.
\item The cones $S_x^\gamma$ are invariant with respect to action of
$\matr{DT}$, and $C_x^\gamma$ are invariant with respect to action
of $\matr{DT}^{-1}$, i.e., for all $x\in A$
$\matr{DT}_x(S_x^\gamma)\subset S_{\matr{T}(x)}^\gamma$ and
$\matr{DT}_x^{-1}(C_x^\gamma)\subset C_{\matr{T}^{-1}(x)}^\gamma$.
\end{enumerate}
\end{theorem}

If the formulated conditions are valid for all points of the
absorbing domain $U$, they are obviously true for the attractor
$A\subset U$. Therefore, the following procedure can be performed
for a verification of the conditions of the theorem.

Starting at $\vec{x}=\{\Re a, \Im a, \Re b, \Im b\}\in U$, we
solve Eqs.~\eqref{eAmplEqns} numerically on the interval
$t\in[0,T]$ and get the image $\vec{x}^\prime=\matr{T}(\vec{x})$.
Also, we initialize four sets of equations for small
perturbations~\eqref{eAmplEqnsVariat} with unit vectors
$\{1,0,0,0\}$, $\{0,1,0,0\}$, $\{0,0,1,0\}$ and $\{0,0,0,1\}$,
respectively, and solve these equations simultaneously with the
original system. From the resulting four vector-columns we compose
a matrix $\matr{U}=\matr{DT}_x$.

If the Poincar\'{e} map is iterated one time from $\vec{x}$, any
perturbation vector $\vec{u}$ transforms to
$\vec{u}^\prime=\matr{U}\vec{u}$. A squared Euclidean norm of this
vector is $||\vec{u}^\prime||^2=\vec{u}^\mathrm{T}
\matr{U}^\mathrm{T}\matr{U}\vec{u}$ where $\mathrm{T}$ means the
transposition. Via the inverse matrix $\matr{U}^{-1}$ we can write
$\vec{u}=\matr{U}^{-1} \vec{u}^\prime$ and
$||\vec{u}||^2=\vec{u}^{\prime\,\mathrm{T}}
\matr{U}^{-1,\mathrm{T}} \matr{U}^{-1}\vec{u}^\prime$. A condition
that the preimage of $\vec{u}^\prime$ relates to the expanding
cone $S_x^\gamma$ is an inequality $||\vec{u}^\prime||\geq\gamma
||\vec{u}||$, or
\begin{equation}
  \label{eConeCond1}
  \vec{u}^{\prime\,\mathrm{T}} \matr{H}_\gamma \vec{u}^\prime\leq 0,
\end{equation}
where
$\matr{H}_\gamma=\matr{U}^{-1,\mathrm{T}}\matr{U}^{-1}-\gamma^{-2}$.

If we start from $\vec{x}^\prime=\matr{T}(\vec{x})$, a vector
$\vec{u}^\prime$ transforms to
$\vec{u}^{\prime\prime}=\matr{U}^\prime\vec{u}^\prime$, and we
have
$||\vec{u}^{\prime\prime}||^2=\vec{u}^{\prime\,\mathrm{T}}\matr{U}^{\prime\,\mathrm{T}}
\matr{U}^\prime\vec{u}^\prime$. The expanding cone
$S_{\matr{T}(x)}^\gamma$ at $\vec{x}^\prime=\matr{T}(\vec{x})$ is
determined by an inequality $||\vec{u}^{\prime\prime}||\geq\gamma
||\vec{u}^\prime||$, or
\begin{equation}
  \label{eConeCond2}
  \vec{u}^{\prime\,\mathrm{T}}\matr{H}_\gamma^\prime \vec{u}^\prime\geq 0,
\end{equation}
where
$\matr{H}_\gamma^\prime=\matr{U}^{\prime\,\mathrm{T}}\matr{U}^\prime-\gamma^2$.

Thus, the required condition $\matr{DT}_x(S_x^\gamma)\subset
S_{\matr{T}(x)}^\gamma$ is formulated in terms of two quadratic
forms: If the inequality~\eqref{eConeCond1} holds, then the
inequality~\eqref{eConeCond2} must be valid too.

Let us perform a canonical reduction of the quadratic form
$\vec{u}^{\prime\,\mathrm{T}}\matr{H}_\gamma^\prime
\vec{u}^\prime$ by a coordinate change. Because the matrix
$\matr{U}^{\prime\,\mathrm{T}}\matr{U}^\prime$ is real and
symmetric, an orthonormal basis of eigenvectors $\vec{d}_0$,
$\vec{d}_1$, $\vec{d}_2$, $\vec{d}_3$ may be chosen. Then, the
matrix $\matr{D}=\{\vec{d}_0, \vec{d}_1, \vec{d}_2, \vec{d}_3\}$
is a diagonalizer:
$\matr{D}^\mathrm{T}\matr{U}^{\prime\,\mathrm{T}}\matr{U}^\prime\matr{D}=
\{\Lambda_i^2\delta_{ij}\}$, $i,j=0,1,2,3$. The eigenvalues on the
diagonal $\Lambda_i^2$ are supposed to be arranged in the
decreasing order. In our case there will be one stretching and
three contracting directions, so, $\Lambda_0^2>1$,
$\Lambda_{1,2,3}^2<1$. Let $\gamma$ be selected in such a way that
$\Lambda_0^2>\gamma^2$, $\Lambda_{1,2,3}^2<\gamma^2$. Under the
transformation $\matr{D}$ the matrix $\matr{H}_\gamma^\prime$ also
becomes a diagonal:
$$\matr{D}^\mathrm{T}
\matr{H}_\gamma^\prime\matr{D}=\matr{D}^\mathrm{T}(\matr{U}^{\prime\,\mathrm{T}}\matr{U}^\prime-
\gamma^2)\matr{D}=\{(\Lambda_i^2-\gamma^2)\delta_{ij}\};$$ here
one diagonal element is positive and others are
negative.\footnote{This property is naturally checked in the
course of computations at each analyzed point of the absorbing
domain: its violation would entail an incorrect operation of
taking a square root of a negative number. The inequalities for
eigenvalues of the matrix $\matr{U}_x^\mathrm{T}\matr{U}_x$ ensure
fulfilment of the condition that a sum of subsets of the linear
vector space (that is a set of all possible linear combinations of
vectors from the expanding and contracting cones) is the full 4D
vector space: $S_x^\gamma+C_x^\gamma=\mathbb{V}_x$.} By additional
dilatation (compression) $\matr{S}=\{s_i^{-1}\delta_{ij}\}$,
$s_0=\sqrt{\Lambda_0^2-\gamma^2}$,
$s_{1,2,3}=\sqrt{\gamma^2-\Lambda_{1,2,3}^2}$, we get
$$\tilde{\matr{H}}_\gamma^\prime=\matr{S}^\mathrm{T}\matr{D}^\mathrm{T}
(\matr{U}^{\prime\,\mathrm{T}}\matr{U}^\prime-\gamma^2)\matr{D}\matr{S}=\sigma_i\delta_{ij},$$
$\sigma_0=1$, $\sigma_{1,2,3}=-1$. The same transformation applied
to the matrix
$\matr{H}_\gamma=\matr{U}^{-1,\mathrm{T}}\matr{U}^{-1}-\gamma^{-2}$,
yields
$$\tilde{\matr{H}}_\gamma=\matr{S}^\mathrm{T}\matr{D}^\mathrm{T}
(\matr{U}^{-1,\mathrm{T}}\matr{U}^{-1}-\gamma^{-2})\matr{D}\matr{S}=
\{h_{ij}\},$$ where $h_{ij}=h_{ji}$.

A condition~\eqref{eConeCond2} for vector
$\vec{c}=\{1,c_1,c_2,c_3\}$ to belong to the expanding cone
$S_{\matr{T}(x)}^\gamma$ now reads as
$c^\mathrm{T}\tilde{\matr{H}}_\gamma^\prime\vec{c}\geq 0$, or
$$c_1^2+c_2^2+c_3^2\leq 1.$$ In the 3D space $\{c_1,c_2,c_3\}$
this corresponds to the interior of the unit ball. A
condition~\eqref{eConeCond1} that preimage of the vector
$\vec{c}=\{1,c_1,c_2,c_3\}$ belongs to the expanding cone
$S_x^\gamma$ is
$\vec{c}^\mathrm{T}\tilde{\matr{H}}_\gamma\vec{c}\leq 0$ or
$$h_{00}+ \sum_{\alpha=1}^3(h_{0\alpha}c_\alpha+h_{\alpha
0}c_\alpha)+ \sum_{\alpha,\beta=1}^3 h_{\alpha\beta}c_\alpha
c_\beta\leq 0.$$ In the space $\{c_1,c_2,c_3\}$ this relation
determines the interior of a certain ellipsoid.

The inclusion $\matr{DT}_x(S_x^\gamma)\subset
S_{\matr{T}(x)}^\gamma$ will be fulfilled, if the ellipsoid is
placed inside the unit ball. Let us formulate an inequality
sufficient to ensure such a disposition. We can evaluate
coordinates for the center of the ellipsoid from a set of linear
algebraic equations
\begin{equation}
  \sum_{\beta=1}^3 h_{\alpha\beta}\bar{c}_\beta = -h_{\alpha 0},\;
  \alpha=1,2,3,
\end{equation}
and then estimate a distance of this point from the center of the
unit ball:
$$\rho=\sqrt{\bar{c}_1^2+\bar{c}_2^2+\bar{c}_3^2}.$$
After transfer of the origin to the point $\{\bar{c}_1, \bar{c}_2,
\bar{c}_3\}$, the equation for the surface of the ellipsoid
becomes $\sum_{\alpha,\beta=1}^3 h_{\alpha\beta}\tilde{c}_\alpha
\tilde{c}_\beta=R^2$, where
$\tilde{c}_\alpha=c_\alpha-\bar{c}_\alpha$ and
$R^2=-h_{00}-\sum_{\alpha=1}^3
(h_{0\alpha}\bar{c}_\alpha+h_{\alpha
0}\bar{c}_\alpha)-\sum_{\alpha,\beta=1}^3
h_{\alpha\beta}\bar{c}_\alpha\bar{c}_\beta$.

Now consider a symmetric $3\times 3$ matrix
$\matr{h}=\{h_{\alpha\beta}\}$. In the diagonal representation
obtained with an orthogonal coordinates transformation
$(\tilde{c}_1, \tilde{c}_2, \tilde{c}_3)\to (\xi_1, \xi_2, \xi_3)$
the equation for the ellipsoid surface takes a form
$l_1\xi_1^2+l_2\xi_2^2+l_3\xi_3^2=R^2$, where $l_1$, $l_2$, $l_3$
are eigenvalues of the matrix $\matr{h}$. The largest principal
semiaxis of the ellipsoid is expressed via the smallest
eigenvalue: $r_\mathrm{max}=R/\sqrt{l_\mathrm{min}}$. Now, an
obvious sufficient condition for the ellipsoid to be positioned
inside the unit ball is given by the inequality
\begin{equation}
  \label{eElInSphere}
  f=r_\mathrm{max}+\rho<1.
\end{equation}
It completes the procedure of the verification of the expanding
cones inclusion for the point $\vec{x}$.

It may be shown that the above procedure, applied to the points of
the absorbing domain $U$ with $\gamma<1$, is equivalent to the
verification of the condition for the contracting cones in the
domain $\matr{x}\in\matr{T}^2(U)$ with the parameter
$\gamma^\prime=1/\gamma>1$:
$\matr{DT}_x^{-1}(C_x^{1/\gamma})\subset
C_{\matr{T}^{-1}(x)}^{1/\gamma}$. It is so because the cones
$S^\gamma$ and $C^{1/\gamma}$ are complimentary sets:
$\bar{S}^\gamma \cup \bar{C}^{1/\gamma}=\mathbb{V}$. (Here
$S^\gamma$ corresponds to the cone of vectors that either expand
or contract but no stronger than by factor $\gamma$.) Hence,
fulfillment of the inequality~\eqref{eElInSphere} checked inside
$U$ for two parameters $\gamma$ and $1/\gamma$ would imply that
both conditions for expanding and for contracting cones are valid
in the domain $\matr{T}^2(U)$ which contains the
attractor.\footnote{The cones $S^\gamma$ and $C^\gamma$ have a
common border only at $\gamma=1$, while for $\gamma>1$ they do not
intersect, as required by the theorem condition: $S_x^\gamma \cap
C_x^\gamma=\varnothing$.} This is sufficient to draw a conclusion
on the hyperbolic nature of the attractor.

\section{Numerical verification of the hyperbolicity}

Numerical tests that we performed according to the above method
for the Poincar\'{e} map of the system~\eqref{eAmplEqns} confirms
that the theorem is fulfilled. Thus, the chaotic attractor under
our consideration is indeed hyperbolic.

For our computer programs we employed algorithms for computation
of eigenvalues of real symmetric matrix, for matrix inversion and
for solution of a set of linear algebraic equations found
in~\cite{Recipes}.

\begin{figure}
  \centering a)\twofig{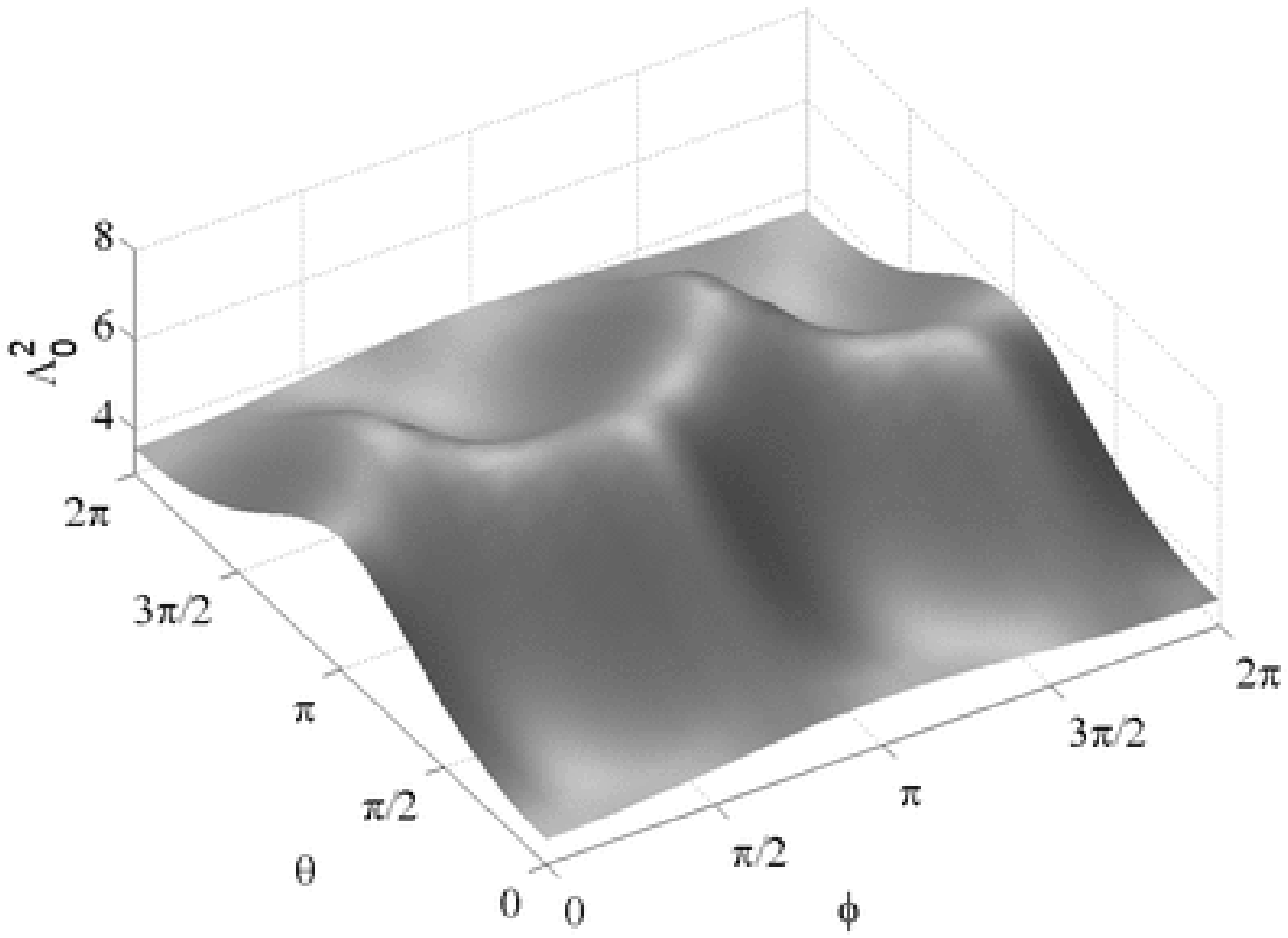}b)\twofig{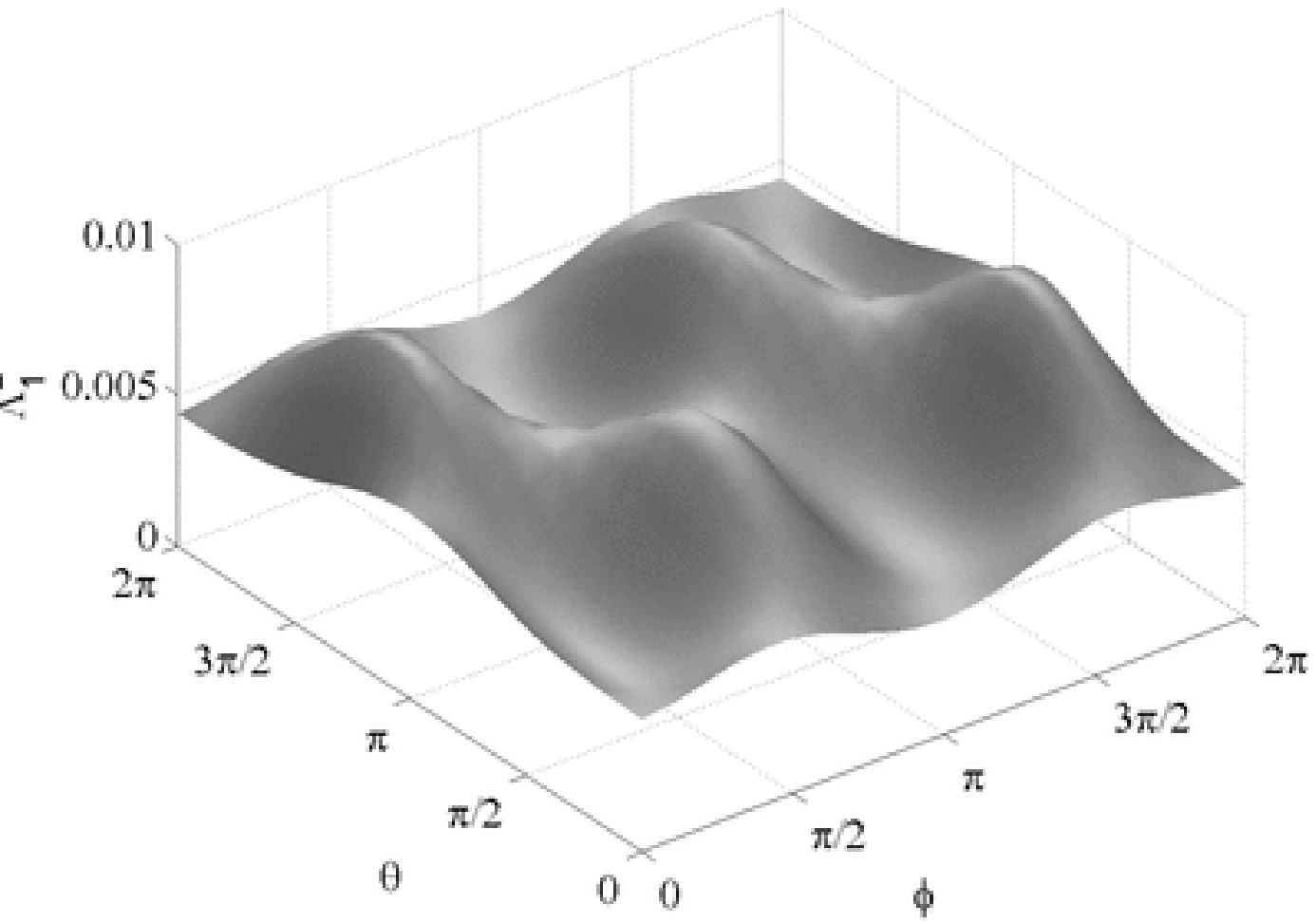}
  \caption{\label{fEigVals} The first and second eigenvalues of
  the matrix
  $\matr{U}_x^\mathrm{T}\matr{U}_x$ vs. two angle variable at the
  boundary of the absorbing area and at $\psi=0$.}
\end{figure}

Necessary conditions, required for the expanding and contracting
cones to be defined, are the inequalities for eigenvalues of the
matrix $\matr{U}_x^\mathrm{T}\matr{U}_x$: $\Lambda_0^2>1$ and
$\Lambda_1^2<1$, respectively. Performed computations demonstrate
that these inequalities are valid in the entire absorbing domain
for the parameters set~\eqref{eNumVal}. The the toroidal absorbing
domain is defined by Eqs.~\eqref{eTorusParam} where $\chi$ and
$\psi$ are kept constant while $\phi$ and $\theta$ are varied from
0 to $2\pi$ with the step $2\pi/50$. At each point we compute the
matrix $\matr{U}_x$ and find corresponding eigenvalues
$\Lambda_i$. Fig.~\ref{fEigVals} represents $\Lambda_0^2$ and
$\Lambda_1^2$ vs. two angle variables at $\chi=1$ (this
corresponds to the surface of the absorbing domain) and at
$\psi=0$. Observe that $\Lambda_0^2>1$ while $\Lambda_1^2<1$ (and
two rest eigenvalues $\Lambda_{2,3}^2$ are also less then one).
Similar graphs appears at other values of $\psi$ and so the same
is for the interior of the absorbing domain.

\begin{figure}
  \bigfig{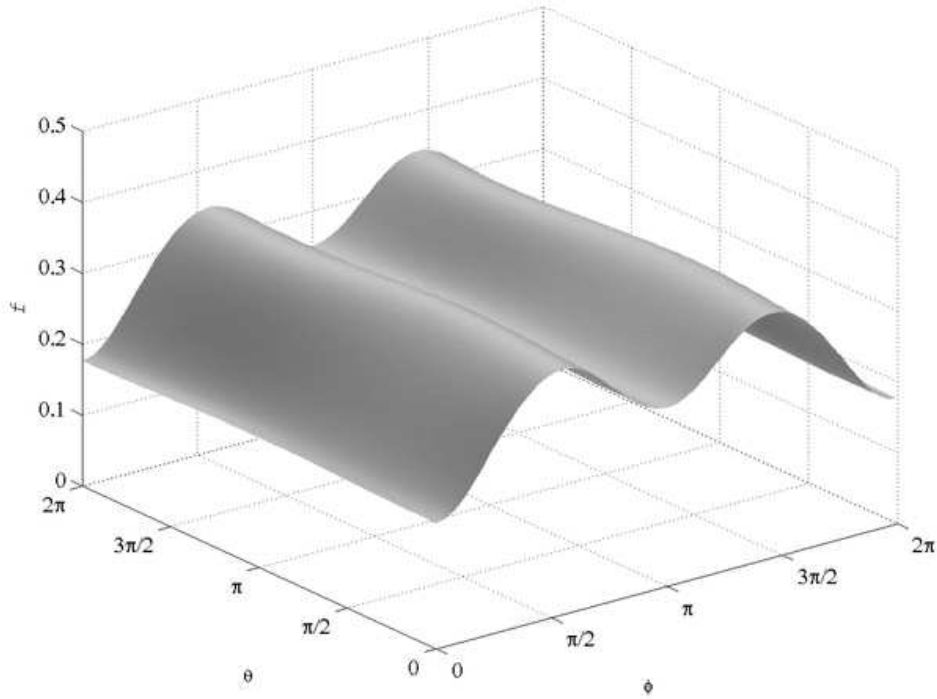}
  \caption{\label{fAngf}Verification of the
  inequality~\eqref{eElInSphere} on the surface of the absorbing
  domain at $\psi=0$, $\gamma=1.1$ and at the parameters values~\eqref{eNumVal}.
  Observe, that $f<1$.}
\end{figure}

Fig.~\ref{fAngf} shows values of the function
$f$~\eqref{eElInSphere} that it takes on the surface of the
absorbing domain at $\chi=1$, $\psi=0$, $\gamma=1.1$ and at the
parameters values~\eqref{eNumVal}. We see that the function $f$ is
always less then $0.5$. The function is smooth and varies
sufficiently slow, so that there are no peaks which can be
candidates for violation of the inequality~\eqref{eElInSphere}
between the checked values of $\phi$ and $\theta$. Thus, we can
conclude that the inequality~\eqref{eElInSphere} is valid.

\begin{figure}
  \centering a)\twofig{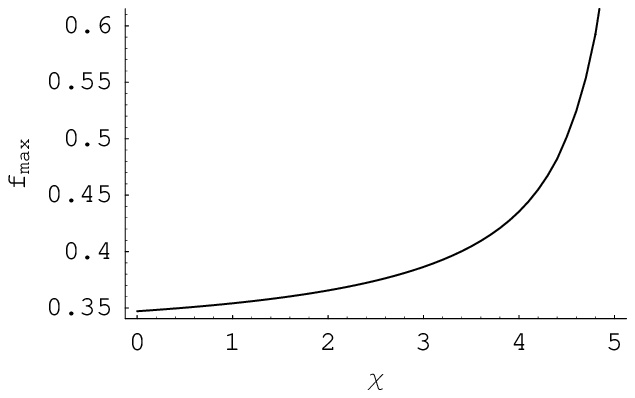}b)\twofig{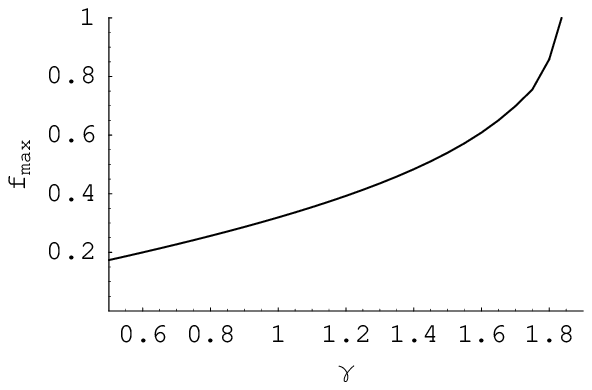}
  \caption{\label{fMaxDist}a) Maximum of $f(\phi,\psi,\theta)$ vs.
  $\chi$ at $\gamma=1.1$. b) Maximum of $f(\phi,\psi,\theta,\chi)$
  vs. $\gamma$. The parameters values are given by~\eqref{eNumVal}.}
\end{figure}

One can see from Fig.~\ref{fAngf} that $f$ depends essentially on
$\phi$ only, while the dependance on $\theta$ is very weak. The
dependance on $\psi$ is tested to be also weak. Plots calculated
at different values of $\psi$ looks almost identical to the that
shown in Fig.~\ref{fAngf}. Qualitatively similar plots appear also
at $\chi<1$ that corresponds to the interior of the absorbing
domain and the inequality~\eqref{eElInSphere} remains always
valid.

To verify the validity of~\eqref{eElInSphere} for the whole
absorbing domain, we fix certain value of $\chi$ and find the
maximum of $f$ as a function of three angle variables $\phi$,
$\psi$, $\theta$. Thus, varying $\chi$, we obtain a function
$f_\mathrm{max}(\chi)$ that is shown in Fig.~\ref{fMaxDist}(a).
Notice that the inequality $f_\mathrm{max}<1$ is valid both inside
the absorbing domain, i.e., at $0\leq\chi \leq 1$, and far beyond.

Now we need to examine the validity of~\eqref{eElInSphere} at
different $\gamma$. We assume that $f$ is a function of four
variables $\phi$, $\psi$, $\theta$ and $\chi$ and find its maximum
at each $\gamma$ with additional requirement $0\leq \chi\leq 1$.
As follows from Fig.~\ref{fMaxDist}(a), $f_\mathrm{max}$ grows
monotonically vs. $\chi$. Therefore, when a numerical procedure,
that seeks a maximum, goes out of the absorbing domain and asks a
value of $f$ at $\chi>1$, we return $f$ at $2-\chi$ instead. In
fact, we create an artificial maximum at the boundary of the
absorbing domain, that prevents the overrun the boundary. If the
seeking procedure can not find a maximum in the interior of the
absorbing domain, it finds the maximum on the boundary at
$\chi=1$. The resulting graph $f_\mathrm{max}(\gamma)$ is shown in
Fig.~\ref{fMaxDist}(b). We see that the inequality
$f_\mathrm{max}<1$ is definitely fulfilled in a wide range of
$\gamma$ both below and above the point $\gamma=1$. This confirms
the expected mutual location of expanding and contracting cones in
the absorbing domain. Hence, we conclude that the hyperbolicity of
the attractor of the system~\eqref{eAmplEqns} is established. This
result can be reproduced for a wide range of values of the
parameters of the system~\eqref{eAmplEqns}.

\section{Conclusion}

In this paper we introduce amplitude equations for a system of two
coupled non-autonomous self-oscillators that have uniformly
hyperbolic chaotic attractor. We provide an evidence that the
system of amplitude equations also demonstrates the hyperbolic
chaotic dynamics. Given a certain set of parameters, we find a
toroidal absorbing domain that is mapped into itself and contains
an attractor of a Poincar\'{e} map for this system. Performed
computations confirm the validity of sufficient conditions for the
hyperbolicity. The conditions are formulated in terms of
inclusions of expanding and contracting cones that are defined in
a tangent vector space associated with the points of the absorbing
domain.

Because of the universality of amplitude equations, the considered
system can be a model for dynamics of various physical systems.
With this example it is possible to construct other models with
hyperbolic chaos, exploiting structural stability of the
hyperbolic attractor. A physical experiment demonstrating
attractor of this type has been performed already on a basis of
coupled electronic oscillators~\cite{HypExper}. In applications,
the systems with hyperbolic chaos may be of special interest
because of their robustness (structural stability). An interesting
and now a substantial direction is constructing chains, lattices,
networks on a base of elements with hyperbolic
chaos~\cite{BunimSinai88}. Models of this class may be of interest
for understanding deep and fundamental questions, like the problem
of turbulence.

This work was partially supported by DFG and RFBR (Grant No.
06-02-16619).

\bibliographystyle{elsart-num-sort}
\bibliography{coneamp}

\end{document}